# Temporal pattern generation with tunable repetition rate using semiconductor laser periodic dynamics

APOSTOLOS ARGYRIS

*Institute for Cross-Disciplinary Physics and Complex Systems – IFISC (CSIC-UIB), Campus UIB, Ctra. de Valldemossa 7.5km, Palma 07122, Spain*
*apostolos@ifisc.uib-csic.es

**In this work, a long-cavity semiconductor laser subject to optical feedback is exploited to generate repetitive temporal patterns with enhanced intra-pattern sample diversity. Limit cycle dynamics characterized by multiple frequency harmonics are experimentally demonstrated by considering a 2 m external cavity with appropriate optical feedback conditions. This simple configuration can enable stable and continuous generation of periodic waveforms with a tunable repetition rate ranging from 3.28 to 4.21 GHz. Distinct, highly consistent temporal patterns are obtained by applying high-pass filtering to adjust the relative power of the harmonics. The resulting system operates as a physical waveform generator capable of producing diverse and repeatable signal patterns, making it well-suited for use in the masking stage of time delay reservoir computing architectures.**

Semiconductor lasers (SLs) subject to optical feedback exhibit a wide range of nonlinear dynamical behaviors that are of both fundamental and practical interest [1-3]. The interaction between delayed optical feedback and the internal carrier–photon dynamics can destabilize the steady-state emission, giving rise to rich temporal phenomena such as periodic or quasi-periodic oscillations, low-frequency fluctuations (LFF), and fully developed broadband chaos. These dynamical regimes have been studied extensively over the past three decades [4-11], while the emission signals have been exploited in various applications, such as in secure optical communications [12], physical random number generation [13], chaos-based sensing [14], microwave and millimeter frequency tone generation [15], and photonic reservoir computing [16]. In the context of photonic reservoir computing, SLs with optical feedback have been particularly attractive as simplified implementations of optical reservoirs, often employing time-multiplexing schemes [17,18]. However, this architecture typically requires pre-processing of the input data – known as masking – at the input stage to enable efficient computation [19]. The mask consists of a sequence of temporal samples with random or low-correlation amplitudes, repeated at the rate of information encoding. Conventionally, such masks are generated using arbitrary waveform generators (AWGs) [16-18]. An alternative approach to circumvent the use of AWGs involves configuring physical systems capable of producing repetitive temporal patterns with intrinsically low-correlation samples. Lately, semiconductor VCSELs with delayed orthogonal polarization feedback were used to generate repetitive square oscillating patterns [20] and mode-locked ultrashort optical pulses [21]. Another promising method, proposed in [22], uses an SL with optical feedback designed to operate in a periodic regime, characterized by the presence of multiple frequency harmonics. In that study, numerical simulations demonstrated that a semiconductor laser with an ultra-short (centimeter-scale) external cavity can exhibit low-complexity instabilities, including periodic temporal patterns.

In this work, I experimentally demonstrate similar dynamical behavior by combining a long-cavity SL with a 2m external optical cavity, extending the findings of the prior numerical study. The experimental setup, illustrated in Fig. 1, employs a straightforward design. The optical feedback into the laser is regulated by an optical attenuator, while polarization stability throughout the system is maintained using polarization-maintaining single-mode fiber. The optical emission is detected by a 40GHz photoreceiver (Thorlabs RX40AF), generating a photocurrent that serves as the microwave electrical signal used to produce repetitive patterns. When specified, a digital RF high-pass filter (HPF) is applied to the photocurrent to provide a spectrally filtered output.

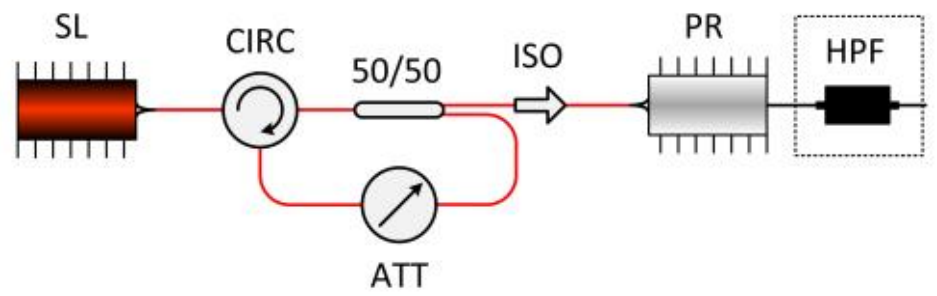

Fig. 1. Experimental setup of a semiconductor laser with external tunable optical feedback. The length of the fiber-based external cavity is 2m. CIRC: 3-port circulator, 50/50: optical splitter, ISO: optical isolator, ATT: optical attenuator, PR: photoreceiver, The HPF is applied to obtain the temporal patterns with increased diversity.

In this configuration, several parameters critically influence the resulting dynamical behavior. Among the most significant and controllable are the laser’s bias current and the strength of the re-injected optical feedback. Additionally, the intrinsic properties of the SL itself play a crucial role in determining the complexity of the optical emission. Standard distributed feedback (DFB) or Discrete Mode (DMLD) laser diodes, which are commonly accessible for telecom applications, are optimized for low threshold current and high slope efficiency. These devices normally feature short internal cavity lengths, usually between 200 and 400 μm. But when such devices are integrated into long external cavities – as in the configuration shown in Fig. 1 – they tend to exhibit predominantly chaotic dynamics. However, operating in a chaotic dynamical regime results to a signal that is unpredictable, thus not repetitive. Consequently, chaotic signals cannot be used as sources for generating repetitive patterns. Still, periodic or quasi-periodic emission regimes can be observed under specific optical feedback conditions. A way to expand the accessible parameter space for such dynamical behavior is to consider SLs with significantly larger internal cavities. Fortunately, certain domains and applications – such as coherent optical communications, high-resolution spectroscopy, and LIDAR – demand SLs with exceptionally narrow linewidths, much lower than the MHz-scale that telecom SLs offer. Achieving such linewidth scales requires careful engineering of the laser device, which may include reducing the linewidth enhancement factor, minimizing internal cavity losses, or increasing the cavity length. Since the laser linewidth is determined by the photon lifetime in the cavity and the resonator's quality factor, a longer cavity increases the photon round-trip time and decreases the optical field decay rate, thereby narrowing the spectral linewidth [23-25]. This relationship is quantitatively described by the modified Schawlow–Townes linewidth equation [25]. By considering such long SL cavities and in presence of strong optical feedback, limit cycle dynamics and multiple frequency harmonics can be observed under extended operating conditions.

In the present study, I incorporate a low-linewidth (100 kHz) distributed feedback laser diode (DMLD), fabricated by Eblana Photonics, in the external cavity configuration shown in Fig. 1. The length of the DMLD is measured to be 1.5 mm, using the side modes of the optical emission spectrum (measured with an Anritsu MS9740B optical spectrum analyzer). The threshold bias current ($I_{th}$) is measured to be 32 mA. By tuning $I_{th}$, I investigate the resulting microwave spectra using a Keysight EXA N9010B RF spectrum analyzer. During this investigation, various levels of optical attenuation were tested, revealing that strong coupling with the external cavity can induce limit cycle dynamics over a broad range of bias currents. The results presented here correspond to a configuration with 3 dB optical attenuation in the feedback path, yielding an overall optical power feedback ratio of 0.12. While higher attenuation levels – corresponding to weaker optical feedback – also permit the observation of limit cycle dynamics, this occurs over a narrower range of DMLD bias currents. Notably, this behavior is not observed in SLs with short internal cavities under the same feedback configuration. Specifically, by replacing in the setup of Fig. 1 the low-linewidth DMLD with a DFB laser (300 μm cavity length) or a conventional telecom DMLD (340 μm cavity length) only LFF and chaotic dynamics are observed.

Fig. 2 presents the experimentally obtained RF emission spectra for increasing laser bias currents of the low-linewidth DMLD. Slightly above $I_{th}$, external cavity modes (ECMs) emerge around the laser's relaxation oscillation frequency, with a mode spacing defined by the external cavity length (52 MHz), resulting in a multi-frequency emission (Fig. 2a). As the bias current increases – and consequently the relaxation oscillation frequency of the SL rises – the supported ECMs become confined around this higher frequency and are progressively filtered out from the system's dynamical response (Fig. 2b). Eventually, for bias currents between $1.48{\cdot}I_{th}$ and $1.71{\cdot}I_{th}$, the ECMs are strongly suppressed, resulting in a spectrum of high-order, equidistant, harmonic frequencies (Fig. 2c, 2d). This behavior can be considered a self-pulsation operation, associated to coherent oscillations due to the nonlinear mixing of the feedback and the relaxation dynamics of the SL. Beyond $1.71{\cdot}I_{th}$, the system enters into a continuous wave emission. The corresponding optical spectra are included in the Supplementary Material (Section 1). The fundamental frequency, along with its harmonics, increases linearly with the laser bias current, as illustrated in Fig. 3. The fundamental frequency spans from 3.28 GHz to 4.21 GHz, providing nearly 1 GHz of tunability in the emitted temporal pattern, while maintaining consistent dynamical behavior. The higher-order peaks are integer multiples of the fundamental frequency. As the bias current increases, the relaxation frequency continues to shift to higher values, causing a corresponding shift in the harmonic frequencies. In parallel, the power of the highest-order harmonics reduces until they disappear in the noise floor. As a result, while up to five harmonics contribute to the limit cycle dynamics in the low bias current regime, only the first three are prominent at higher bias levels (Fig. 3).

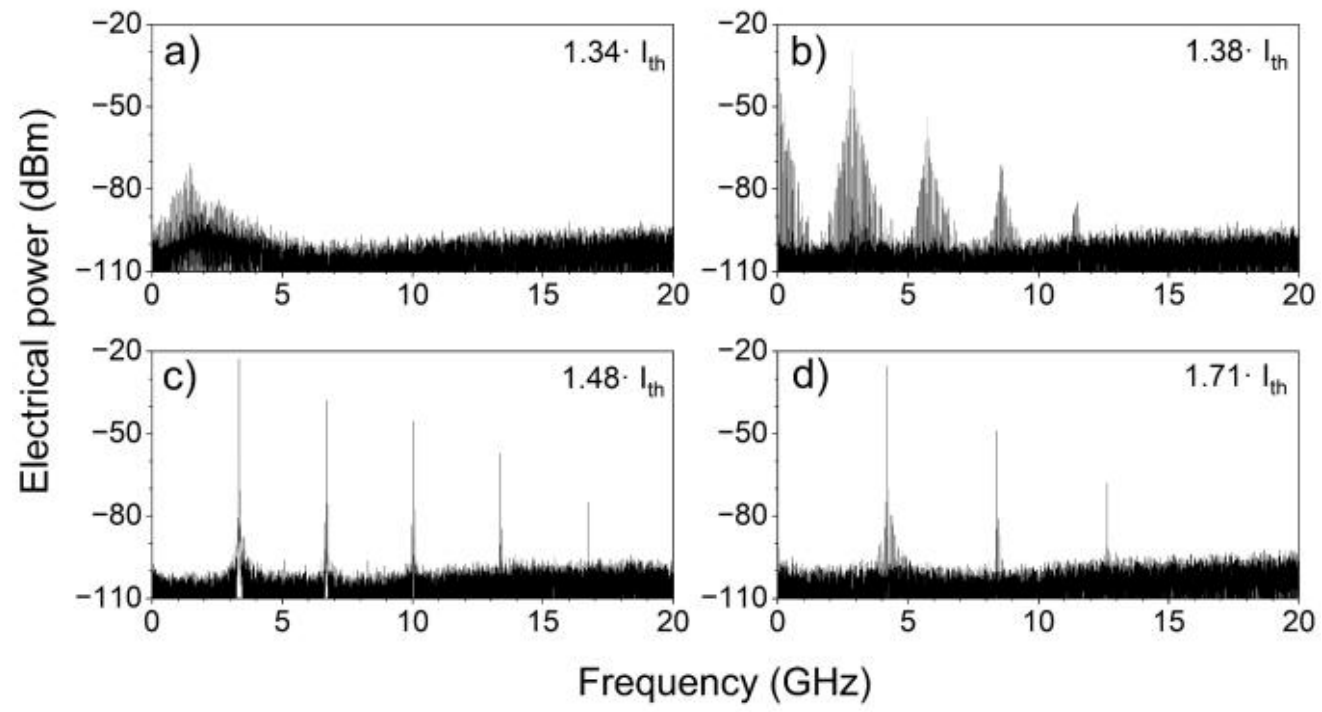

Fig. 2. Experimentally obtained RF emission spectra after photodetection, for different SL's bias currents: a) $1.34{\cdot}I_{th}$, b) $1.38{\cdot}I_{th}$, c) $1.48{\cdot}I_{th}$, and d) $1.71{\cdot}I_{th}$.

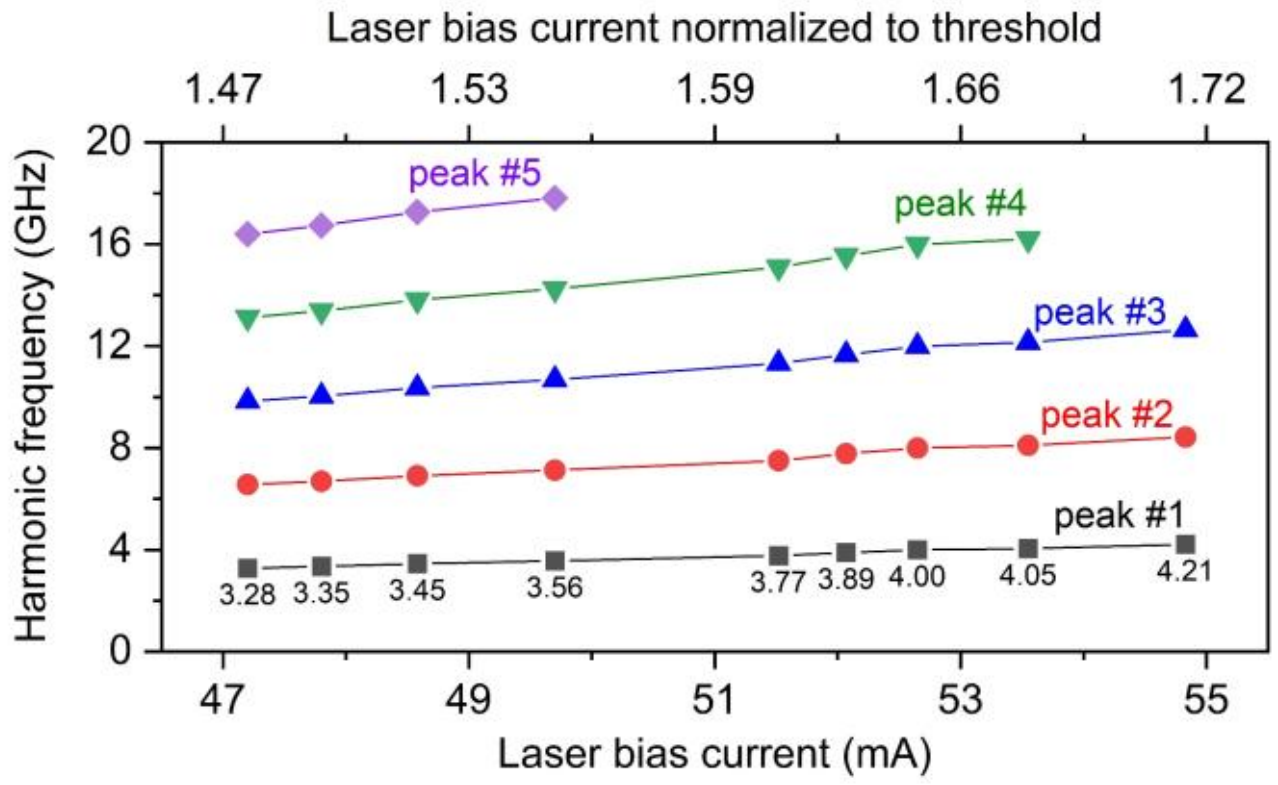

Fig. 3. Frequency of the harmonic peaks of the periodic emission versus the SL's bias current.

Figure 4a illustrates a representative temporal waveform of the periodic emission from the SL at a bias current of 47.24 mA ($1.48{\cdot}I_{th}$). The corresponding limit cycle attractor, reconstructed using time-delay embedding, is shown in Fig. 4b. The time delay for attractor unfolding was determined as the time at which the autocorrelation function decays to $e^{-1}$ of its maximum value. At this bias current, the fundamental frequency which defines the repetition rate is 3.28 GHz, corresponding to a quasi-sinusoidal period of 305 ps. Other operating conditions could be also used to slightly tune the shape of the original pattern. The acquired waveform samples (marked as red dots) have a temporal resolution of 3.9 ps, and emerge from the 256 GSa/s sampling rate of the monitoring oscilloscope. Within this periodic waveform, a high degree of intra-pattern correlation is observed among the samples. The primary objective, however, is to generate an analog temporal waveform in which the effective samples exhibit minimal mutual correlation. Such a waveform would be well-suited as a masking sequence for time-delay reservoir computing. As demonstrated in prior studies, greater diversity in the amplitudes of the masking samples enhances the dimensionality expansion in time-delay reservoir systems, thereby improving performance on computational tasks.

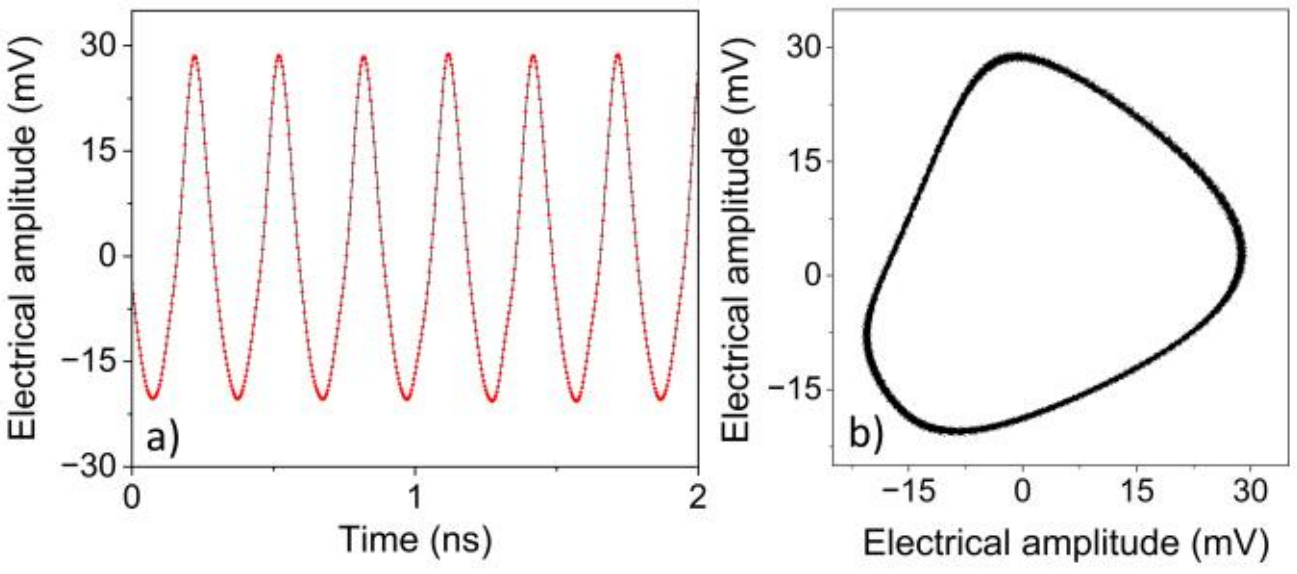

Fig. 4. a) Temporal repetitive pattern sampled at 256 GSa/s, and b) attractor of the periodic emission of the SL for a bias current of 47.24 mA ($1.48{\cdot}I_{th}$).

An analysis of the power spectral distribution and relative intensity of the harmonic components (Figs. 2b–d) reveals a gradual attenuation of the higher-order harmonics. This smoothing arises from several interrelated physical mechanisms that act as filters or dampers for the high-frequency components. First, the finite carrier lifetime in the semiconductor medium imposes a temporal response limit, acting as a low-pass filter. Rapid variations in the optical field, associated with higher-frequency harmonics, are not efficiently supported by the gain medium. Second, the gain spectrum of the laser has a finite bandwidth, favoring amplification near the relaxation oscillation frequency. Additionally, the laser cavity itself supports a discrete set of longitudinal modes determined by its length and effective refractive index. These cavity resonances function as a comb filter, selectively reinforcing certain frequencies while attenuating others. Higher-order harmonics, being more distant from resonance peaks, couple less efficiently into the cavity and thus exhibit reduced intensity. Collectively, these filtering effects attenuate high-frequency components in the emission spectrum, which correspond to sharp transitions in the time-domain waveform. As a result, the output signal becomes smoother in time and exhibits the observed spectral roll-off. This outcome is contrary to our objective.

To enhance the contribution of higher-order harmonics and reduce the dominance of the fundamental frequency, a 5th-order high-pass Butterworth electrical filter (HPF) is applied after photodetection (Fig. 1). By varying the low-cutoff frequency of the HPF, the shape of the limit cycle attractor – and consequently the temporal pattern – changes significantly, while the overall duration of the repetitive pattern remains unchanged. I consider the example case of the $1.48{\cdot}I_{th}$ bias current, that corresponds to the original pattern of Fig. 4, and apply HPFs with different low-cutoff frequencies at 4, 5, 6 and 7 GHz. The resulting attractors and time-domain waveforms are presented in Figs. 5 and 6, respectively. From the RF spectrum in Fig. 2c, it is evident that the first three filters (with 4, 5, and 6 GHz cutoff frequency) progressively attenuate the fundamental frequency. In contrast, the 7 GHz HPF also attenuates the second harmonic, thereby altering the spectral content and the resulting waveform more drastically.

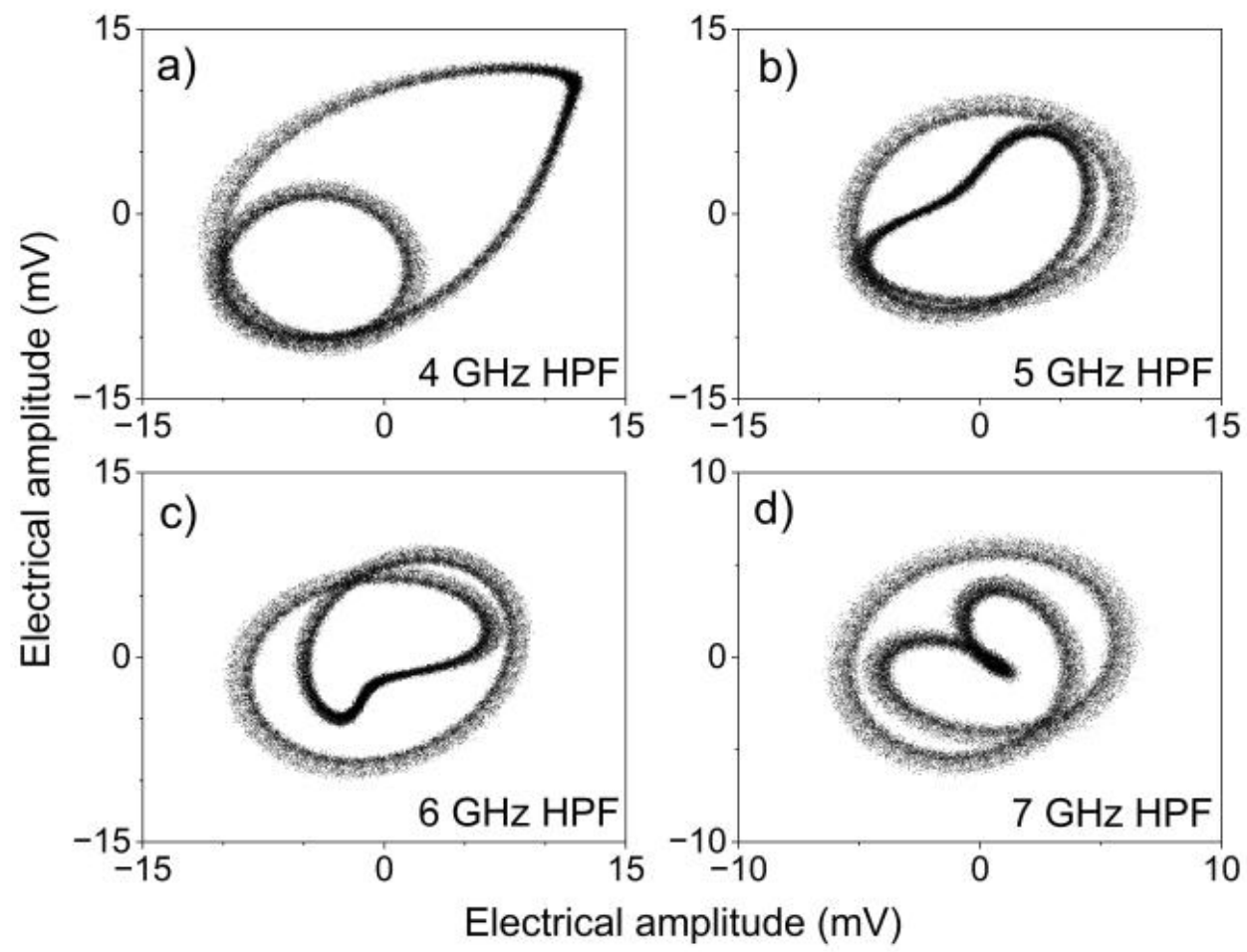

Fig. 5. Attractors of high-pass filtered time series for a SL's bias current of 47.24 mA ($1.48{\cdot}I_{th}$) and for different cut-off frequencies: a) 4 GHz, b) 5 GHz, c) 6 GHz, and d) 7 GHz.

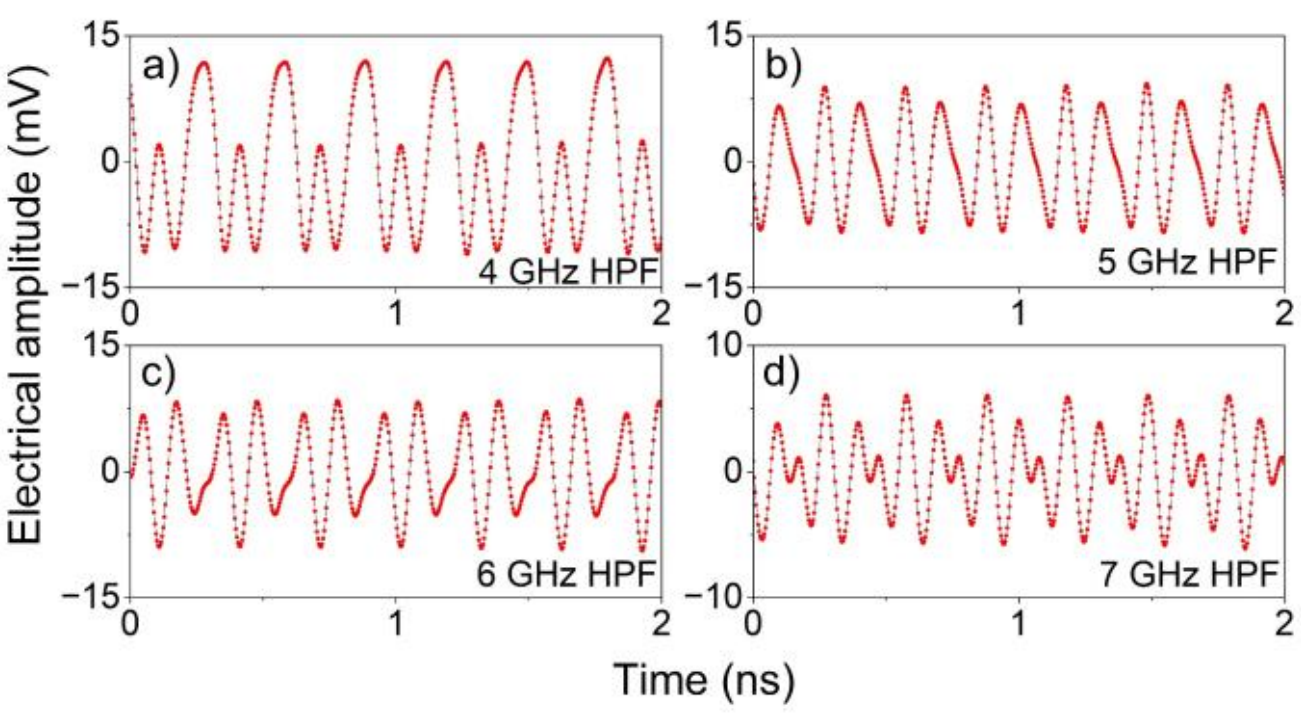

Fig. 6. Temporal repetitive patterns that correspond to the attractors of Fig. 5, and are obtained from high-pass filtered time series for a SL's bias current of 47.24 mA ($1.48{\cdot}I_{th}$) and for different cut-off frequencies: a) 4 GHz, b) 5 GHz, c) 6 GHz, and d) 7 GHz.

A preliminary visual inspection reveals that the quasi-sinusoidal temporal pattern of the original, unfiltered signal has been transformed into patterns with increased intra-pattern diversity. This diversity can be quantitatively assessed using various statistical tools. An effective and computationally efficient metric for evaluating the diversity of temporal samples in a repetitive time series, particularly when adjacent samples exhibit strong correlations, is the Effective Sample Size (ESS). ESS estimates the number of statistically independent samples within the series and is derived from the autocorrelation function. It is computed as:

$$ESS = \frac{N}{1 + 2\sum_{k=1}^{m} \rho_k} \tag{1}$$

with the following boundaries: $1 \leq ESS \leq N$, where $N$ is the total number of samples within the pattern. A higher ESS indicates greater diversity among the samples, implying reduced temporal redundancy. $\rho_k$ represents the autocorrelation at lag $k$, evaluated up to a maximum lag $M$, beyond which the autocorrelation becomes negligible. This method explicitly accounts for temporal dependencies in the data, making it particularly suitable for high sampling rate acquisition scenarios where strong autocorrelations between adjacent samples prevail.

The ESS is employed here to evaluate the sample diversity of the original and high-pass filtered temporal patterns, as shown in Fig. 7. When considering all experimentally acquired samples within a pattern (black rectangles), the filtered patterns exhibit a higher ESS than the unfiltered one, with the 5 GHz HPF yielding the highest ESS value. One approach to further reduce sample correlation is to apply over-sampling - i.e., selecting every $m$-th sample from the complete set. As $m$ increases, the number of available samples within a single pattern period decreases, but the remaining samples exhibit reduced correlation. This is shown in Fig. 7 for values of $m$ ranging from 2 to 5 (indicated by different colors). In all cases, the high-pass filtered patterns consistently exhibit greater diversity than the original signal. Depending on the down-sampling scale, the optimized cut-off frequency of the HPF can be lower or higher.

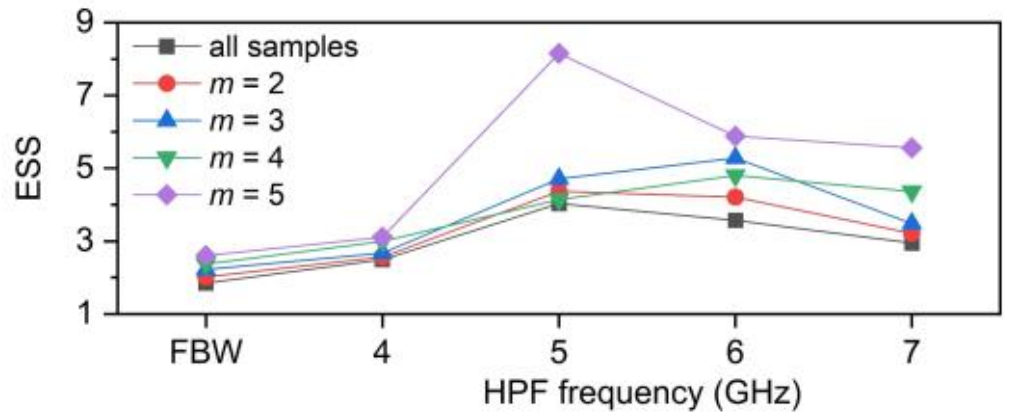

Fig. 7. ESS for the original (full bandwidth - FBW) and the high-pass filtered patterns, for a laser bias current of 47.24 mA (1.48·$I_{th}$). Down-sampling of the pattern results in different ESS values.

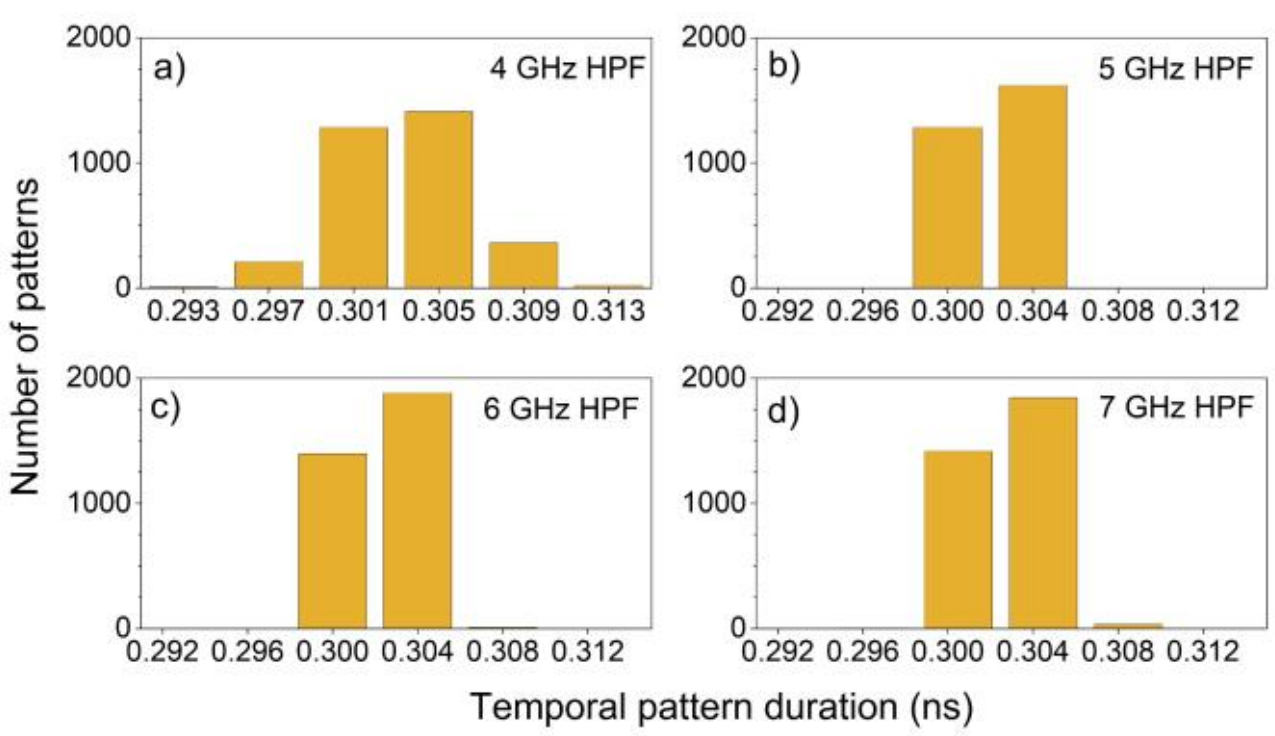

Fig. 8. Histograms of the duration of the temporal repetitive patterns obtained from HPF time series for a SL's bias current of 47.24 mA (1.48·$I_{th}$) and for different cut-off frequencies: a) 4 GHz, b) 5 GHz, c) 6 GHz, and d) 7 GHz. The total duration of the analyzed time series is 1μs.

A final aspect that is worth evaluating is the temporal consistency of the generated patterns over timescales significantly longer than the pattern duration. To this end, a statistical analysis of the pattern duration across extended recordings is performed. Specifically, 1 μs-long time series of the photodetected signal are acquired under the previously examined HPF conditions. The resulting histograms of pattern duration are presented in Fig. 8. Given the oscilloscope's sampling resolution of 3.9 ps, the temporal resolution in identifying the duration of each pattern is ±3.9 ps. Among the cases studied, only the 4 GHz HPF condition shows a slight deviation in the pattern duration that exceeds the sampling resolution. For the higher cutoff frequencies (Figs. 8c–d), the measured pattern durations are fully determined by the sampling resolution, indicating highly consistent pattern generation.

In conclusion, I have experimentally demonstrated that a SL with a long internal cavity and external optical feedback can serve as a stable and tunable source of multiple harmonic frequency tones. By applying appropriate electrical high-pass filtering after photodetection, a variety of repetitive temporal patterns can be generated, with a tunable rate spanning nearly a 1 GHz range. This system offers a physically implemented masking pre-processing technique suitable for time delay reservoir computing. It has also all the attributes to be considered in a photonic integrated setting, with short external optical cavity (Supplementary Material, section 2).

**Acknowledgment.** This work was supported by the Program for Centres and Units of Excellence in R&D María de Maeztu under the project CEX2021-001164-M, and the project INFOLANET PID2022-139409NB-I00, both funded by MCIN/AEI/10.13039/501100011033.

**Disclosures.** The author declares no conflicts of interest.

**Data Availability.** Data underlying the results presented in this paper are not publicly available at this time but may be obtained from the authors upon reasonable request.

## Full references